# Forecasting residential heating and electricity demand with scalable, high-resolution, open-source models

Stephen J. Lee [a,b,*], Cailinn Drouin [a]

[a] MIT Energy Initiative, Massachusetts Institute of Technology, United States of America

[b] Manning College of Information & Computer Sciences, University of Massachusetts Amherst, United States of America

## HIGHLIGHTS

- Building-level, hourly heating and electricity demand forecasting.
- 18.8% and 27.6% RMSE reductions vs baseline for heating and electricity demand.
- 59% WIS reduction for both heating and electricity demand.
- Heat pump classification via comparing estimates to consumption (ROC-AUC 0.77).
- Additional heat pump classification using building attributes alone (ROC-AUC 0.72).

## GRAPHICAL ABSTRACT

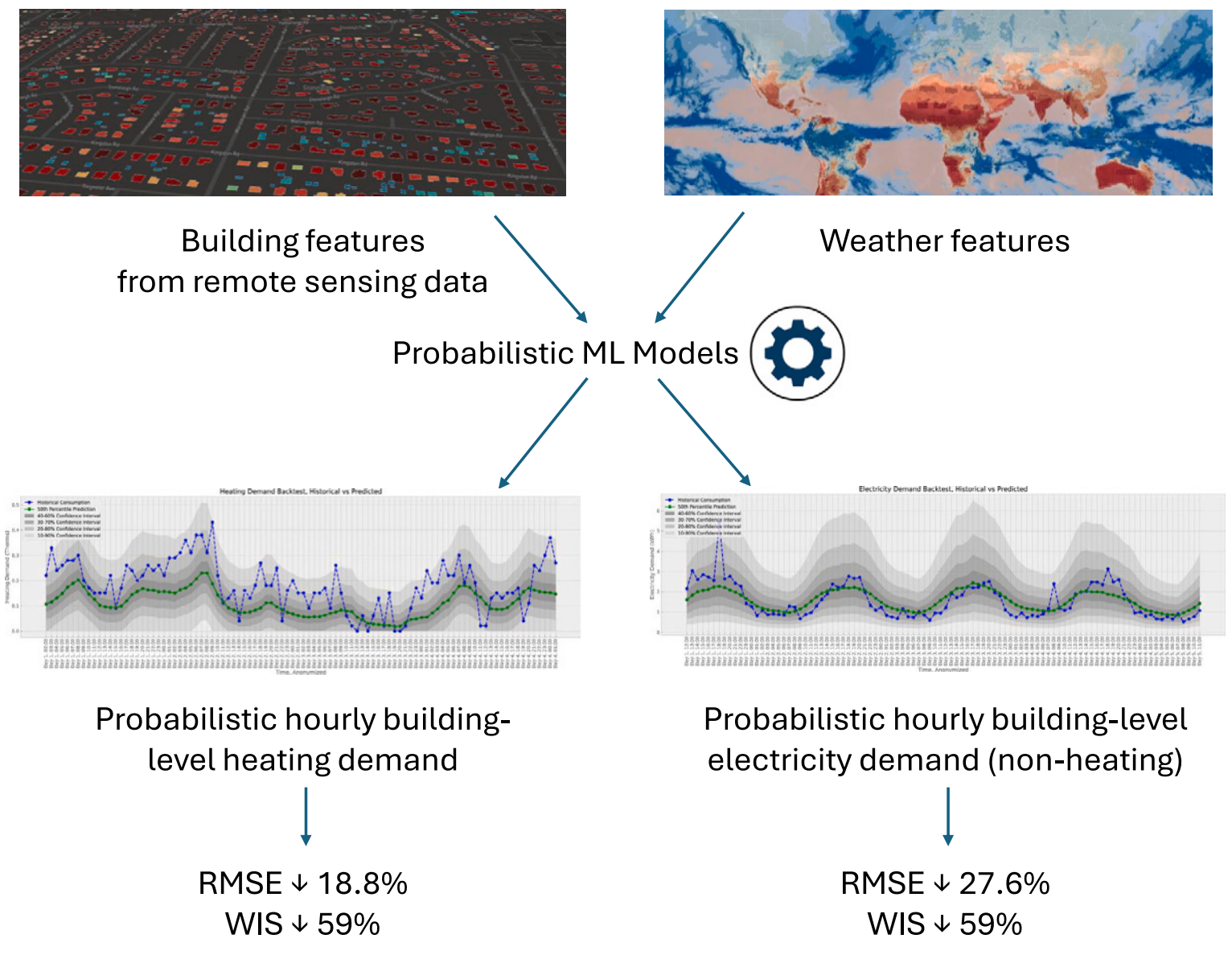

## ARTICLE INFO

## ABSTRACT

Electrifying space and water heating is a critical priority for the energy transition. The necessary widespread adoption of heat pumps will have significant impacts on the power grid. Studies report heating electrification may increase winter peak electricity demand by up to 70%, with some colder regions experiencing a more than fourfold increase in peak demand. Contending with this increased demand will necessitate unprecedented upgrades to the power grid.

The process of upgrading the grid has a critical spatial dimension, as heating demand, electricity demand, and the capacity of existing grid infrastructure vary significantly across regions. Grid planning also involves a critical temporal dimension: short-term weather patterns and long-term climate change introduce complexities and uncertainties that can be difficult to quantify. However, most existing demand forecasts are provided and validated only at aggregated spatial scales, lack temporal detail, and provide single-valued predictions. Without accurate, probabilistic, and spatially and temporally resolved demand forecasts, planners risk misallocating scarce resources.

* Corresponding author at: Manning College of Information & Computer Sciences, University of Massachusetts Amherst, United States of America.
*E-mail address:* stephenl@umass.edu (S.J. Lee).

We present a novel framework for high-resolution forecasting of residential heating demand and non-heating electricity demand using probabilistic deep learning models. Because our models are trained on electricity consumption from a predominantly gas-heated region, the learned electricity demand patterns primarily reflect non-heating end uses such as lighting, appliances, and cooling. We focus specifically on providing hourly building-level electricity and heating demand forecasts for the residential sector. Leveraging multimodal building-level information – including data on building footprint areas, heights, nearby building density, nearby building size, land use patterns, and high-resolution weather data – and probabilistic modeling, our methods provide granular insights into demand heterogeneity. Validation at the building level underscores a step change improvement in performance relative to NREL's ResStock model, which has emerged as a research community standard for residential heating and electricity demand characterization. In building-level heating and electricity estimation backtests, our probabilistic models respectively achieve RMSE scores 18.8% and 27.6% lower than those based on ResStock, with probabilistic forecast quality measured via WIS improving by 59% for both applications. By offering an open-source, scalable, high-resolution platform for demand estimation and forecasting, this research advances the tools available for policymakers and grid planners, contributing to the broader effort to decarbonize the U.S. building stock and meeting climate objectives.

## 1. Introduction

Electrifying space and water heating is a critical priority for the energy transition [1,2]. Residential and commercial buildings make up 13% of direct U.S. emissions [3], with fossil-fueled space heating representing the single greatest constituent of this share [4]. Heat pumps offer a promising solution to decarbonizing building heating. By enabling the efficient electrification of heating systems, they provide a pathway for significant emissions reductions as the power grid transitions to cleaner energy sources.

The necessary widespread adoption of heat pumps will have significant impacts on the power grid. For instance, Waite and Modi estimate that full electrification of heating with current high-efficiency heat pump technologies could increase national peak electricity loads by 70%, with some colder regions experiencing a more than fourfold increase in peak demand [5]. Contending with this increased demand will necessitate unprecedented upgrades to the power grid.

The process of upgrading the grid is inherently spatial in nature as heating demand, electricity demand, and the capacity of existing grid infrastructure varies widely across regions. Unfortunately, the vast majority of existing demand forecasts are provided and validated only at aggregated scales, obfuscating the heterogeneity and distribution of demand within communities. They also rarely provide forecasts accounting for short- and long-term trends. Without accurate, spatially and temporally resolved demand forecasts, planners risk critically misallocating scarce resources.

Detailed demand forecasting is essential to address these challenges. By incorporating high levels of spatial and temporal granularity, advanced forecasting models can identify demand at arbitrary levels of spatial and temporal aggregation, estimate localized peak heating and electricity loads, anticipate seasonal variations, predict long-term trends, characterize forecasting uncertainty, and support grid planners optimizing asset investments. This level of detail helps to minimize costs, reduce inefficiencies, and enhance resilience as the grid adapts to support widespread electrification.

We present a novel framework for high-resolution residential heating demand and non-heating electricity demand[1] forecasting employing probabilistic deep learning models. We focus specifically on providing hourly building-level electricity and heating demand forecasts for the residential sector. Leveraging multimodal building-level information – including data on building footprints, heights, nearby building density, land use patterns, and high-resolution weather data – and probabilistic modeling, our methods provide granular insights into demand heterogeneity. Validation at the building level underscores a step change improvement in performance relative to NREL's ResStock model, which has emerged as a research community standard for residential heating and electricity demand characterization. By offering an open-source, scalable, high-resolution platform for demand estimation and forecasting, this research advances the tools available for policymakers and grid planners, contributing to the broader effort to decarbonize the U.S. building stock and meet climate objectives.

We make three contributions:

1. We introduce an open-source, scalable framework that produces hourly, building-level probabilistic forecasts of residential heating and electricity demand using widely available geospatial and weather data.
2. We demonstrate state-of-the-art accuracy at the building level in the U.S. Mid-Atlantic, reducing RMSE relative to NREL's ResStock by 18.8% for heating and 27.6% for electricity, with probabilistic forecast quality measured via weighted interval score (WIS) improving by 59% for both applications.
3. We demonstrate that our probabilistic demand models enable heat pump classification, achieving ROC-AUC of 0.77 using predicted CDF features and 0.72 using remote sensing and census features alone. Unlike existing smart meter-based detection methods, our building attribute classifier enables building-level propensity estimation without requiring historical consumption data.

## 2. Related work

Various modeling approaches have been developed to estimate heating and electricity demand and assess the impacts of electrification. We categorize existing approaches into four types: physics-based simulation models, spatially aggregated demand models, heating degree day regression models, and machine learning models. More detail about related work and individual studies are provided in Supplementary Material Section S2.

**Physics-based simulation models**, such as NREL's ResStock and ComStock, estimate demand using engineering principles and building archetypes [6,7]. While they produce hourly outputs calibrated against surveys (e.g., RECS, CBECS, etc.), their reliance on synthetic archetypes limits accuracy for individual buildings, validation occurs only at state or city levels, and they do not quantify forecast uncertainty [8,9].

**Spatially aggregated models** estimate demand at census tract or county scales using public datasets [10,11]. These support macro-level electrification planning but lack building-level granularity, provide deterministic outputs, and cannot adapt to real-time conditions.

**Heating degree day (HDD) models** offer a parsimonious, temperature-driven approach with minimal data requirements, making them valuable for global-scale analyses and data-sparse regions [12].

[1] Machine learning models for heating demand are learned from historical gas consumption data in a predominantly gas-heated area, while those for non-heating electricity demand come from historical electricity consumption data corresponding to these same areas. For the remainder of this paper, we will refer to 'non-heating electricity demand' simply as 'electricity demand'.

However, they have limited adaptability to building-specific characteristics and do not quantify uncertainty.

**Machine learning models** span diverse applications: short-term forecasting [13–16], annual EUI estimation [17,18], design-phase surrogates [19–22], and long-term grid impact analysis [23]. ML-based natural gas forecasting similarly operates at regional or station levels [24,25], with building-level models often lacking temporal granularity [26–28]. While ML models offer scalability, capture complex patterns, and enable fast inference, they frequently train on simulated data and lack the combined spatial, temporal, and probabilistic resolution needed for building-level infrastructure planning.

**Heat pump penetration and propensity models** address the challenges of identifying existing heat pump installations and predicting future adoption. Detection methods primarily rely on smart meter data: Weigert et al. achieved an AUC of 0.82 for heat pump detection using electricity consumption combined with weather and geographic features [29]. Propensity modeling has identified socio-demographic determinants of adoption, including geography, climate, and housing characteristics [30], with adoption notably consistent across income levels unlike other clean energy technologies [31]. However, smart meter-based detection requires historical consumption data, precluding application to unmetered buildings, while propensity models typically operate at aggregate geographic scales. Extended discussion of heat pump detection and propensity methods is provided in Supplementary Section S2.5.

Our contribution addresses these gaps by: (1) training directly on real-world AMI data rather than simulated outputs; (2) providing hourly, building-level probabilistic forecasts; (3) leveraging remote sensing features for real-world descriptiveness without requiring detailed building specifications; (4) enabling scenario analysis via exogenous weather inputs (short-term forecasts or long-horizon climate projections); (5) jointly forecasting heating and electricity demand for consistent electrification assessment; and (6) demonstrating that learned consumption distributions enable heat pump classification, while remote sensing features alone support propensity estimation without requiring historical consumption records.

## 3. Data

In this section, we provide an overview of the remote sensing features and ground truth labels used for training, validation, and testing our models.

### 3.1. Features

Our machine learning features fall into three primary categories: building attributes, weather variables, and temporal indicators.

To characterize individual buildings in the U.S. Mid-Atlantic region where we have ground truth AMI data, we aggregate and process multiple geospatial datasets. Overture Maps serves as a foundational dataset, providing building polygons and address information [32]. From these polygons, we derive footprint areas and local building density, defined as the number of buildings within a 1 km radius of a given building. Additionally, we compute the mean and standard deviation of nearby building areas to roughly measure the distribution of nearby building sizes. Further, we enhance our dataset by performing spatial joins, linking buildings to additional contextual attributes. Building height data is sourced from the DLR's World Settlement Footprint 3D V2 dataset [33]. Local land use shares for crops, built areas, and rangeland are extracted from Esri's 10-meter Land Cover dataset, aggregated over grids of $1 \times 1$, $11 \times 11$, and $51 \times 51$ pixels [34]. Overhead nighttime light intensity values are derived from the Visible Infrared Imaging Radiometer Suite (VIIRS) and also averaged across $11 \times 11$ and $51 \times 51$ pixel building-centered grids [35]. Additionally, internet connectivity data, including download and upload speeds, latency, and the number of tests for both fixed and mobile devices, are obtained from Ookla's dataset [36]. These features, available at national and even global scales, reflect the potential scalability of our methodology.

To capture socioeconomic characteristics of each building's surrounding area, we incorporate census tract-level variables from the American Community Survey (ACS) 5-Year Estimates (2018–2022) [37]. Specifically, we extract median household income, median year structure built, and housing value indicators including median contract rent, median owner-occupied unit value, and their respective lower and upper quartiles. These variables serve as proxies for building age, construction quality, and occupant demographics, all of which influence heating and electricity consumption patterns.

We source hourly weather data from ERA5, which is produced by the European Centre for Medium-Range Weather Forecasts (ECMWF) under the Copernicus Climate Change Service. ERA5 provides reanalysis climate data at 0.25° spatial resolution from 1940 to the present [38]. Each building is assigned the corresponding ERA5 raster values for the current and each of the two preceding hours for a set of key meteorological variables. These include two-meter air temperature, total precipitation, snow depth, surface pressure, 10-meter wind gusts, zonal/meridional wind components, surface net thermal radiation, surface net solar radiation, and total cloud cover. Collectively, these features impact heating demand, air infiltration, and solar heating gains. Additionally, soil temperature and high vegetation cover are included to account for subsurface heat exchange and the sheltering effects of surrounding vegetation.

To capture cyclical variations in demand, we incorporate a one-hot encoded representation of the day of the week and transform the hour-of-day integer into paired sine and cosine features. This circular encoding preserves a unit distance between successive hours, including the transition from hour 23:00 back to 0:00, preventing exaggerated temporal distances while retaining diurnal structure.

### 3.2. Ground truth labels: AMI data

We obtain ground truth data from natural gas and electricity AMI readings, representing a diverse set of buildings across urban and rural zip codes in the U.S. Mid-Atlantic region. To ensure a representative sample, we draw 60,000 residential buildings per application (heating and electricity): 30,000 for training and 15,000 each for validation and testing. Each building has approximately one year of consecutive hourly consumption data between 2022 and 2023, capturing seasonal variations throughout the year.

The vast majority of space and water heating demand in our training region is met using natural gas. Other end uses of natural gas, including cooking, dryers, pilot lights, fireplaces, and pools, are relatively minor and intermittent. We therefore treat total natural gas consumption as a proxy for heating demand. To account for these other uses and heating efficiency, we apply an assumed conversion efficiency factor $\eta$ to translate fuel input into delivered heat for space heating and water.

We describe this methodology in detail in Section 4.4. To characterize our sample while preserving consumer anonymity, we present building attribute distributions for randomly sampled subsets of the training, validation, and test sets in Supplementary Material Section S3.1.

### 3.3. Ground truth labels: Heat pump rebates

We also obtain an incomplete dataset of buildings with heat pumps derived from local utility rebate programs. This dataset captures buildings that received heat pump rebates during a tracked period, though it is not comprehensive. Buildings absent from this dataset may still have heat pumps installed prior to or after the rebate tracking period began. We lack visibility into whether rebate recipients have hybrid heating systems, full heat pump systems without gas backup, or whether the heat pumps remain operational. Despite these limitations, we treat a positive rebate label as a strong signal that the building has a heat pump installed. To avoid data leakage between heating demand estimation and heat pump classification tasks, we exclude rebate linked buildings from the gas consumption training set used for our heating demand models. Section 4.5 describes how we leverage this rebate dataset for heat pump classification experiments.

### *3.4. Data processing*

To join AMI consumption data with building information from the Overture Maps dataset, we implement a method that standardizes address components and employs fuzzy string matching. Initially, we normalize street names by standardizing common suffixes to ensure consistency. We then group addresses by ZIP code and identify exact matches between street names in the AMI data and the Overture dataset. For street names without exact matches, we compute Levenshtein distance-based normalized similarity scores and identify highly probable matches [39]. We empirically determined a normalized similarity score threshold of 88. String matches with scores greater or equal to 88 are treated as matches while those with scores below this threshold are discarded. Once street names are matched, we further refine the process by matching house numbers to establish precise correspondences between the datasets. Since we use buildings as our unit of analysis from a remote sensing standpoint, we aggregate consumption for all potential downstream apartment units to the concatenated street number and street name level.

## 4. Methods

We present two new distinct models for our applications: one for estimating heating demand and another for electricity demand. They are each comprised of a multi-layer perceptron (MLP) encoding transformations of input features into parameters of the Zero-Inflated Gamma (ZIG) distribution. We refer our model configuration as MLP-ZIG. In this section, we describe these models and outline how we utilize ResStock output to comprise a baseline. We also compare against two standard machine learning baselines, Gradient Boosting Regression (GBR) and Random Forest (RF), trained with quantile loss functions to produce probabilistic forecasts, as well as a reduced-feature MLP-ZIG variant using only the top 10 SHAP-selected features. Full implementation details, hyperparameter sweeps, and quantile inference procedures for all baselines are provided in Supplement Section S6.

### *4.1. Model architecture, training, and validation*

We develop a probabilistic neural network (NN) to model building-level heating and electricity demand with high spatial and temporal resolution. Our model is based on a multi-layer perceptron (MLP) NN architecture and utilizes an objective function that minimizes the mean negative log-likelihood of the Zero-Inflated Gamma (ZIG) distribution. This distribution effectively captures both zero-heavy and skewed data patterns, as described in the following subsection.

The network comprises five fully connected layers with 20, 25, 30, and 25 neurons, respectively, followed by the output layer. We apply dropout layers after each dense layer with tunable dropout rates for regularization. Training employs batch-based stochastic gradient descent with the Adam optimizer. We employ our validation set to conduct a grid search over dropout rates $\in \{0.0, 0.05, 0.10, 0.20\}$ and L2 regularization strengths $\in \{0, 10^{-4}, 10^{-3}, 10^{-2}\}$, yielding 16 configurations per application, and apply adaptive learning rate reduction for computational efficiency. Full hyperparameter ranges and selected values are reported in Supplement Table S2. We chose our basic model parameters through grid search optimization and common heuristics from similar energy demand forecasting research as described in [40]. We standardize all input features with z-score scaling to improve numerical stability. Model evaluation employs deterministic metrics RMSE, MAE, and the Weighted Interval Score (WIS), a proper scoring rule that jointly assesses calibration and sharpness of probabilistic forecasts [41], computed on held-out test data. Mean WIS denotes the average WIS across all test set building-hour observations; lower values indicate sharper and better-calibrated probabilistic forecasts. The full WIS definition is provided in Supplement Section S4.1.

We do not include lagged target variables (past electricity or heating demand). Instead, we provide autoregressive exogenous features, namely ERA5 meteorological variables at times t, t−1, and t−2, together with temporal indicators. At inference, the model produces one-hour-ahead, building-level forecasts at an hourly cadence. Because inputs are exogenous, forecasts can be rolled forward for any horizon with available weather projections, including short-term numerical weather forecasts or long-horizon CMIP6 scenarios.

#### *4.1.1. ZIG likelihood objective function*

In our MLP-ZIG model configuration, we employ the Zero-Inflated Gamma distribution to capture key aspects of heating and electricity demand, including frequent zero values and positively skewed nonzero consumption. The likelihood function consists of a Bernoulli component modeling zero occurrences and a Gamma distribution for positive values:

$$f(x \mid p, k, \theta) = \begin{cases} p, & x = 0, \\ (1-p) \cdot \frac{x^{k-1} e^{-x/\theta}}{\theta^k \Gamma(k)}, & x > 0, \end{cases} \tag{1}$$

where $p$ is the probability of zero consumption, and $k$ and $\theta$ are the shape and scale parameters of the gamma distribution. We train our models using an objective function comprised of the mean negative log-likelihood (NLL) of the ZIG distribution. Within the objective function, $p$ is constrained using the sigmoid function, and $k$ and $\theta$ using the softplus.

#### *4.1.2. Model checking*

We assess calibration via the probability integral transform (PIT): for each test observation, we evaluate its cumulative probability under the predicted distribution. If the model is well-calibrated, these PIT values are uniformly distributed on $[0, 1]$, so their empirical CDF should lie along the diagonal $y = x$. We quantify calibration with the Kolmogorov–Smirnov (KS) statistic, which measures the maximum deviation from uniformity.

#### *4.1.3. Feature importance*

We evaluate feature importance using SHapley Additive exPlanations (SHAP), which provide game-theoretic attributions quantifying each feature's marginal contribution to predictions [42]. For our NNs, we employ DeepExplainer [43], which leverages the compositional structure of deep networks to efficiently estimate Shapley values. We explain the predicted ZIG mean rather than raw logits, ensuring attributions reflect expected consumption. We generate two complementary visualizations: (i) global importance bar plots ranking features by mean absolute SHAP value, and (ii) beeswarm plots displaying the distribution and direction of individual attributions with feature values color-coded. Full methodological details, including background sampling, explainer configuration, and limitations, are provided in Supplement Section S8.1.

### *4.2. Designing a ResStock baseline*

We benchmark our probabilistic models against physics-based simulations derived from ResStock v3.3.0. ResStock was designed to characterize aggregate demand across representative building archetypes rather than to predict consumption at individual premises; nevertheless, it serves as the most widely used reference for residential energy demand and provides a meaningful benchmark for building-level evaluation. To align ResStock's synthetic archetypes with real-world buildings in our study area, we implement a three-stage matching procedure. First, we generate over 13,000 representative archetypes across 36 Public Use Microdata Areas (PUMAs) spanning our Mid-Atlantic study region, each simulated with location-appropriate weather profiles. Second, we spatially join each observed building to its corresponding PUMA and restrict candidate archetypes at the highest resolution available. Third, we estimate building story counts from World Settlement Footprint 3D heights [33], filter archetypes by story, and select the archetype whose total floor area most closely matches the observed building. This procedure ensures that each test building receives demand estimates from a physically plausible archetype while maintaining broad coverage. Full implementation details, including the split-conformal procedure used to derive probabilistic forecasts from the deterministic simulator, are provided in Supplement Section S6.3.

### 4.3. MLP-ZIG reduced features baseline

To assess whether a smaller, more interpretable feature subset suffices, we train a reduced-feature MLP-ZIG variant using only the 10 most informative features as identified by SHAP analysis. This baseline uses identical architecture and training procedures but restricts inputs to top-ranked features, which differ between heating and electricity. Full feature lists and implementation details are provided in Supplement Section S6.1.

### 4.4. Relating natural gas to heating demand

To estimate heating demand, we use AMI readings of natural gas consumption as a proxy, applying an assumed efficiency factor $\eta$ to convert fuel use into delivered thermal energy. In our study region, natural gas is predominantly used for space and water heating, while other uses, including cooking, dryers, pilot lights, fireplaces, and pools, are relatively minor and intermittent. Thus, total gas consumption provides a sufficiently accurate basis for estimating heating demand at the building level.

We derive $\eta$ using simulation outputs from NREL's ResStock v3.3.0. We filter for residential buildings in our service areas that rely exclusively on natural gas for both space and water heating, excluding any that utilize electricity, fuel oil, propane, wood, coal, or solar thermal systems for heating. For each building, we extract total annual gas consumption and total delivered thermal load for space and water heating.

To estimate the conversion efficiency, we fit a constrained linear regression of the form:

$$\text{Delivered Load} = \eta \times \text{Fuel Use}$$

This zero-intercept model assumes that fuel use directly scales with delivered thermal energy and simplifies the transformation of consumption data into heating load.

Since we do not have visibility into the heating system efficiency of individual homes in our AMI dataset, we apply ResStock's inferred efficiency factor in aggregate.

### 4.5. Classification methods for heat pump penetration and propensity

We leverage the learned representations from our MLP-ZIG electricity demand model to classify buildings by heating system type. Using the heat pump rebate dataset described in Section 3.3, we construct balanced training, validation, and test sets containing 4000, 1800, and 1800 buildings respectively, with equal representation of heat pump and non heat pump buildings in each split.

We evaluate two feature engineering approaches. The first approach extracts predicted CDF values from the trained MLP-ZIG model. For each building hour, we compute the CDF value of observed consumption under the predicted distribution, then average these values within 10 equal width temperature bins spanning the observed range. This produces a 10 dimensional feature vector per building that captures how consumption percentile varies with outdoor temperature. The second approach uses only static building attributes from remote sensing and census data, enabling classification without historical consumption records.

We train MLP, random forest, logistic regression, support vector machine classifiers on each feature set. Model selection uses validation set ROC-AUC scores.

Full experimental details are provided in Supplement Section S9.

## 5. Results

In this section, we present results for both heating and electricity demand forecasting, including baseline comparisons, model checking, and feature importance analyses. All results reported correspond to test data.

For the heating demand application, all analysis results are expressed in terms of natural gas consumption. As detailed in Section 4.4, we also present a method to approximate space and water heating demand based on natural gas usage, with corresponding results provided in Section 5.5. While natural gas demand and heating demand are not strictly equivalent, we use the term "heating demand" throughout this section for ease of communication.

Fig. 1 provides qualitative backtest visualizations, illustrating the probabilistic model outputs against historical consumption across four winter and summer days for each application. A single anonymized consumer is reflected in the heating demand backtests in Figs. 1(a) and 1(b), and a different single anonymized consumer for the electricity demand backtests in Figs. 1(c) and 1(d).

### 5.1. Baseline comparisons

We compare the performance of our MLP-ZIG models against the ResStock baseline in Tables 1 and 2 and Fig. 2. These tables present deterministic and probabilistic accuracy metrics. RMSE and median MAE evaluate point forecast accuracy, while WIS evaluates the full predictive distribution. We deliberately omit metrics such as Mean Absolute Log Error and MAPE because the prevalence of zero consumption hours makes those statistics degenerate for single-point evaluation. WIS is our primary probabilistic metric because it is directly computable from the common quantile format produced by all models and baselines, including conformalized ResStock outputs. By contrast, CRPS for quantile-only baselines requires reconstructing or numerically approximating a full predictive distribution from quantiles, which introduces additional interpolation assumptions. In this setting, WIS provides the most direct and comparable probabilistic evaluation across methods.

### 5.2. Clustered aggregation experiments

Because utilities and planners often act on measures of demand and consumption over aggregated regions, we extend our evaluation to assess how prediction errors behave under spatial aggregation. We conduct two complementary experiments: *time-aligned aggregation*, where all buildings in a cluster are sampled from the same hour, and *staggered aggregation*, where buildings are sampled across different time periods.

In staggered aggregation, cluster-level MAPE decreases rapidly with aggregation size. At 2500 buildings, MAPE reaches 2.0% for electricity and 1.7% for heating; at 10,000 buildings, MAPE falls to 1.1% and 0.8% respectively; and at 100,000 buildings, MAPE drops below 0.4% for both end uses. In time-aligned aggregation, errors plateau at higher levels: at 2500 buildings, time-aligned MAPE remains at 28.8% for electricity and 24.6% for heating. This divergence between aggregation modes reveals both the promise of our approach for aggregated demand estimation and the critical role of weather data quality in aggregate-level forecasting accuracy. Full methodology and planning implications are provided in Supplementary Section S7.3.

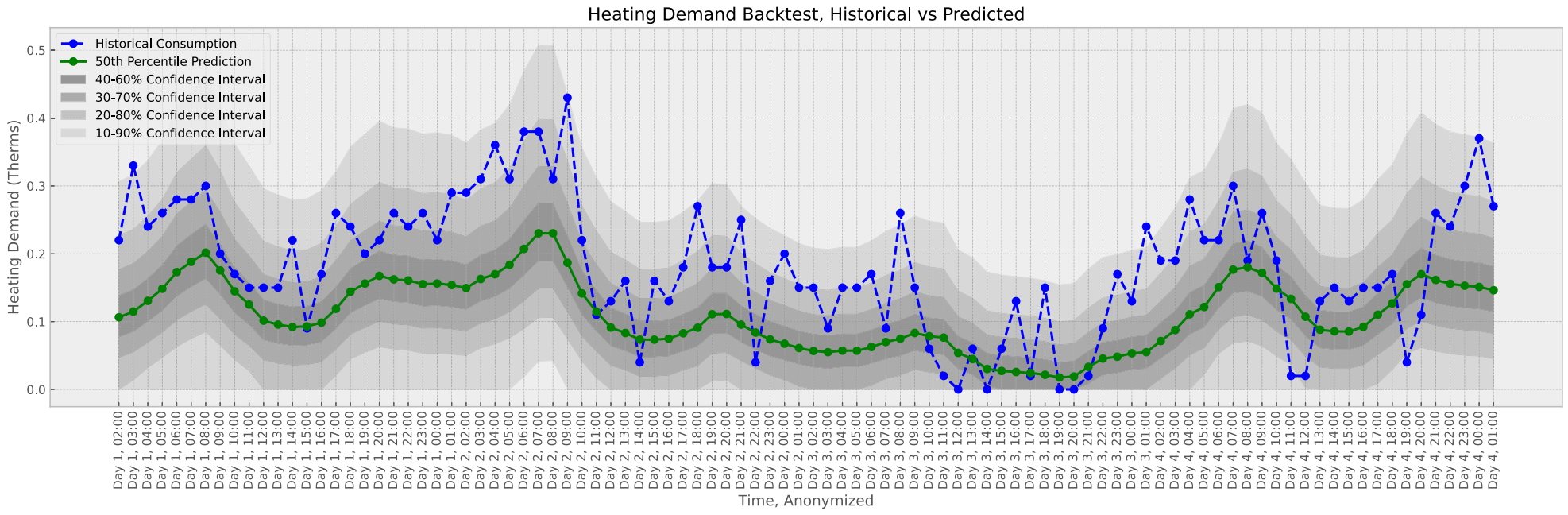

(a) Winter heating demand backtest.

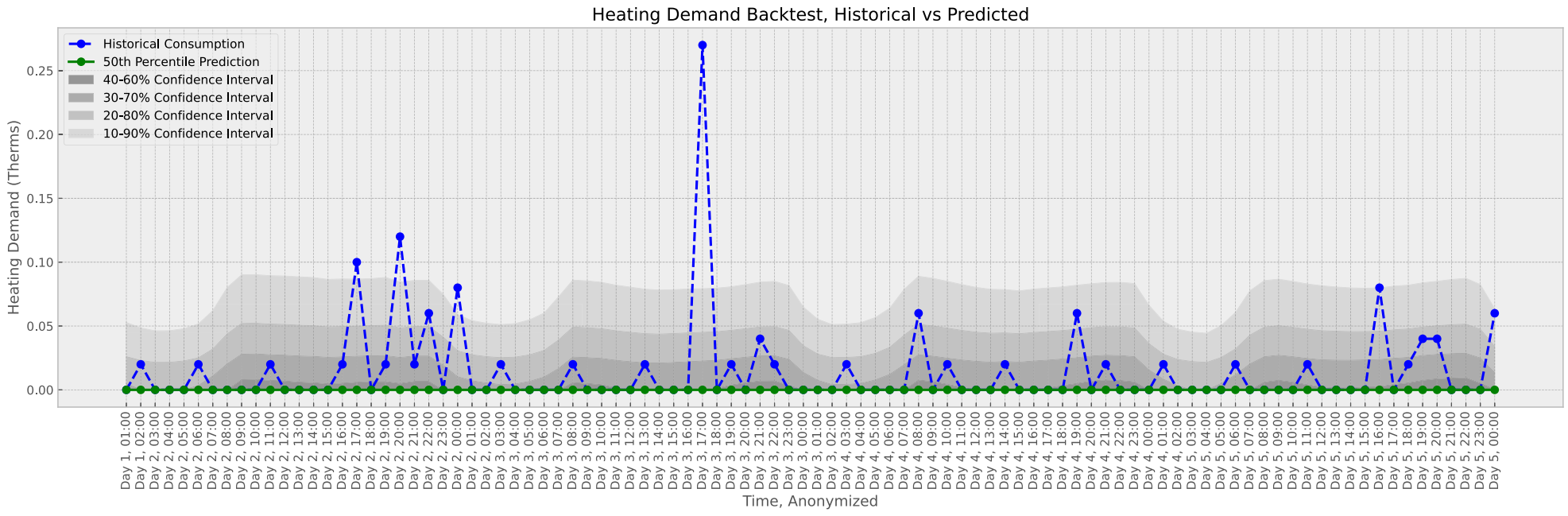

(b) Summer heating demand backtest.

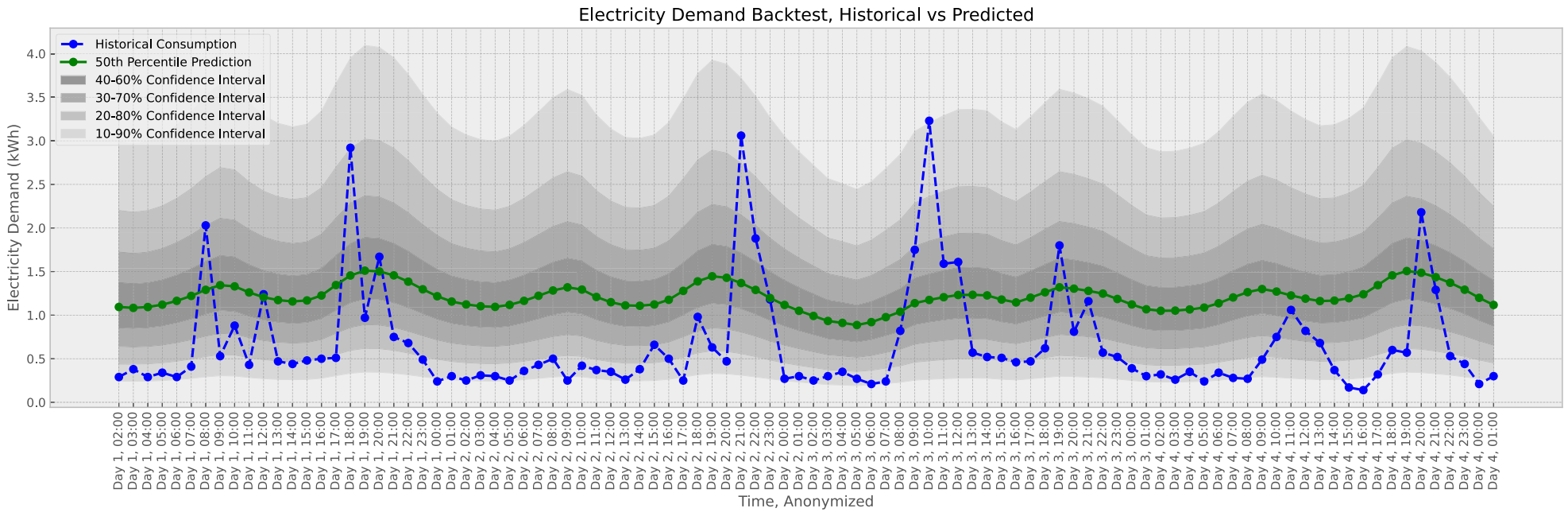

(c) Winter electricity demand backtest.

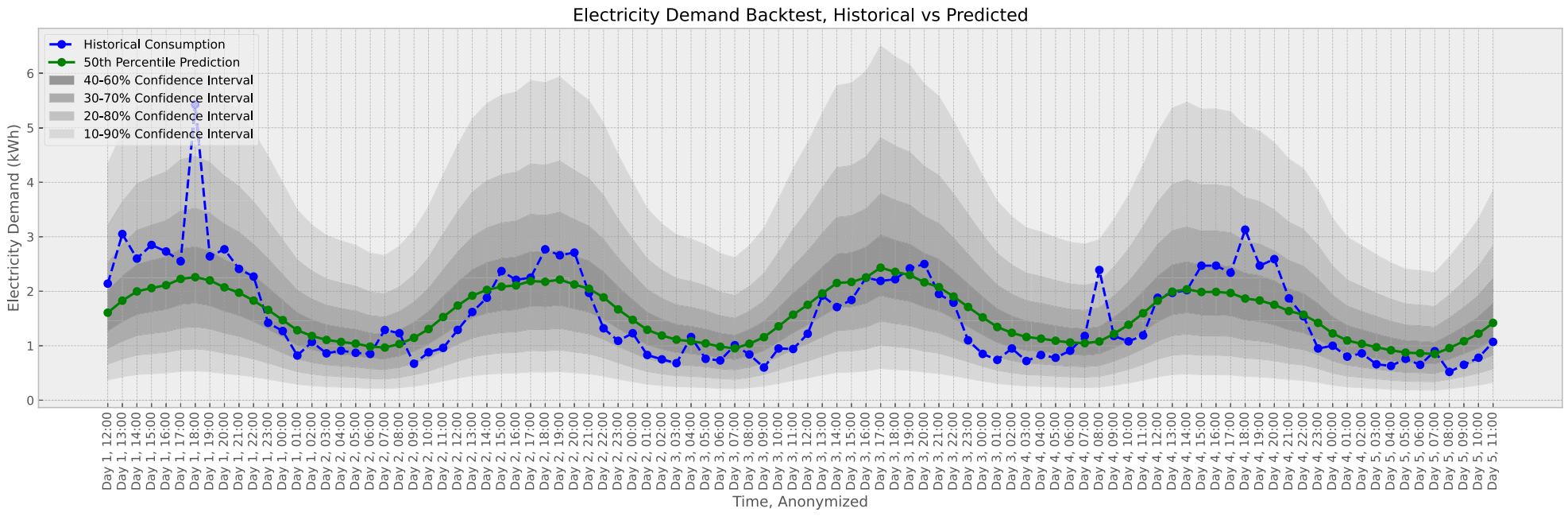

(d) Summer electricity demand backtest.

**Fig. 1.** Seasonal heating and electricity demand backtests comparing MLP-ZIG estimates with historical consumer consumption. The top row presents heating demand results for (a) winter and (b) summer, while the bottom row shows electricity demand results for (c) winter and (d) summer.

### *5.3. Model checking*

We report full calibration diagnostics via empirical CDFs, KS statistics, and interval coverage in Supplementary Material Section S5.2. The electricity model exhibits excellent calibration with a KS statistic of 0.039, indicating that predicted quantiles closely match observed frequencies. Empirical coverage of the 90% prediction interval is 89.5% for the electricity MLP-ZIG model and 89.8% for the limited-feature MLP-ZIG model, indicating close agreement with nominal 90% coverage. The heating model shows larger deviation from uniformity with a KS statistic of 0.285, reflecting the greater challenge of modeling zero-inflated natural gas consumption. For heating, the full MLP-ZIG and limited-feature MLP-ZIG models achieve 90% coverage of 95.5% and 95.1%, respectively, while the benchmark GBR is closest to nominal at 90.7%. Full coverage results at 50%, 80%, 90%, and 95% nominal levels for all models are provided in Supplementary Table S5.

**Table 1**
Comparison of RMSE, MAE, WIS, and 90% interval coverage across electricity demand models. Median values of our predicted distributions are used for point estimates.

| Model | RMSE | Mean absolute error | Mean WIS | 90% Cov. (%) |
|---|---|---|---|---|
| MLP-ZIG | 1.427 | **0.859** | 2.075 | 89.5 |
| MLP-ZIG w/Limited features | **1.419** | **0.859** | **2.067** | **89.8** |
| Benchmark (RF) | 1.454 | 0.862 | 2.191 | 75.8 |
| Benchmark (GBR) | 1.473 | 0.892 | 2.476 | 60.9 |
| ResStock with mapping | 1.960 | 1.195 | 5.092 | 94.1 |

**Table 2**
Comparison of RMSE, MAE, WIS, and 90% interval coverage across heating demand models. Median values of our predicted distributions are used for point estimates.

| Model | RMSE | Mean absolute error | Mean WIS | 90% Cov. (%) |
|---|---|---|---|---|
| MLP-ZIG | **0.108** | **0.053** | **0.122** | 95.5 |
| MLP-ZIG w/Limited features | **0.108** | **0.053** | 0.123 | 95.1 |
| Benchmark (RF) | 0.111 | 0.058 | 0.135 | 89.0 |
| Benchmark (GBR) | 0.112 | 0.059 | 0.136 | **90.7** |
| ResStock with mapping | 0.133 | 0.069 | 0.297 | 94.8 |

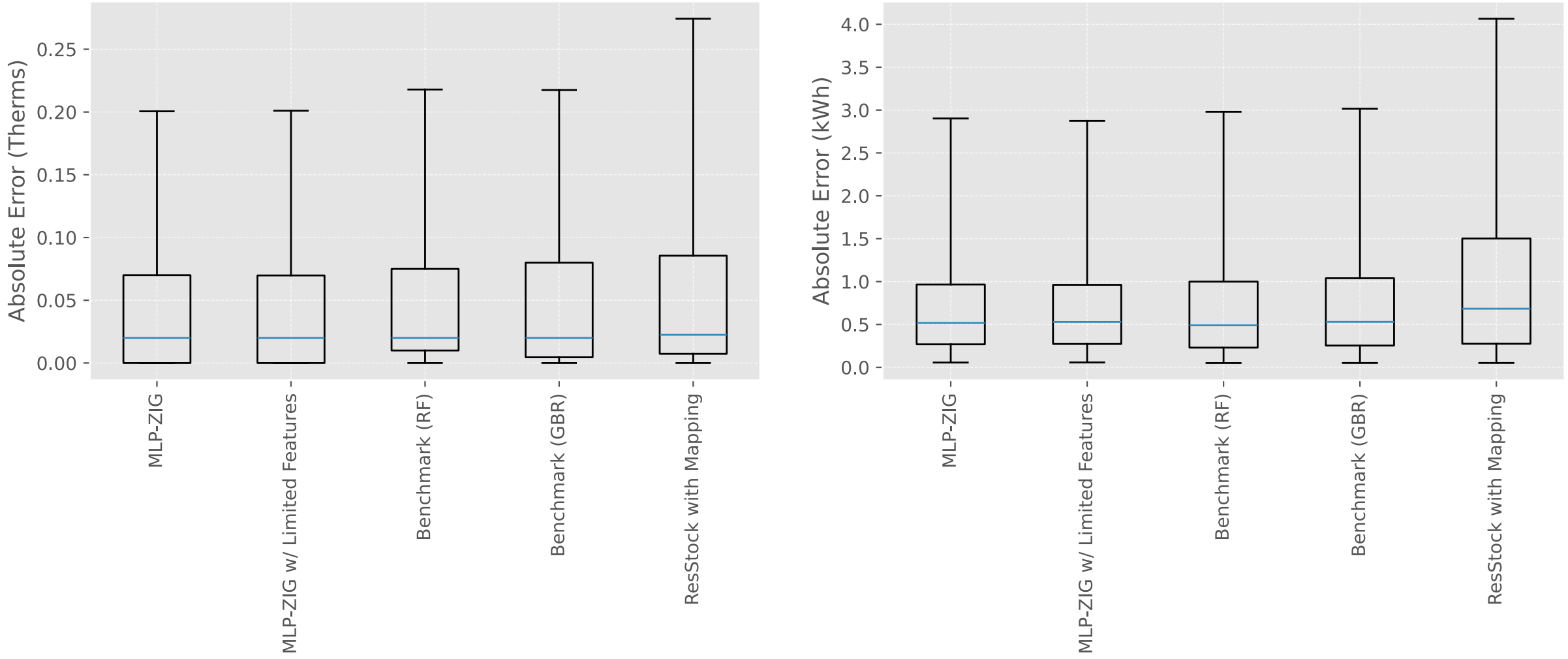

(a) Heating absolute error (therms).

(b) Electricity absolute error (kWh).

**Fig. 2.** Absolute error distributions for heating (therms) and electricity (kWh) across MLP-ZIG, limited-feature, benchmark, and ResStock baselines.

### 5.4. Feature importance

SHAP analysis reveals distinct feature importance profiles for heating and electricity demand. For heating, meteorological features dominate: soil temperature at current and lagged hours, two-meter air temperature at current and lagged hours, building footprint area, temporal indicators, and median year built comprise the top-ranked predictors, reflecting the direct physical relationship between weather and space heating requirements. For electricity, building footprint area emerges as the most influential predictor, followed by temporal indicators, nearby building count, soil and two-meter temperature features at various lags, and land use patterns including tree cover, reflecting the diverse structural and environmental drivers of electricity consumption.

Comprehensive SHAP summary and beeswarm visualizations are provided in Supplementary Material Section S8, including methodology, global importance rankings, and discussion of limitations.

### 5.5. Relating natural gas to heating demand

Applying the method described in Section 4.4 to the filtered ResStock dataset yields an empirical efficiency factor of $\eta = 0.7512$, indicating that, on average, 75.12% of metered gas is converted to end use space and water heating. The regression achieves a coefficient of determination $R^2 = 0.96$, suggesting strong linearity and low variance around the fitted model. Based on ResStock data, this supports the use of a constant efficiency factor for converting gas consumption into thermal demand across diverse building archetypes.

We apply this value of $\eta$ across all buildings in our dataset to transform hourly natural gas estimates into estimated hourly heating demand. Despite simplifications, this method captures the dominant portion of heating-related gas use in our areas of analysis.

### 5.6. Heat pump classification and propensity modeling

We observe that buildings with heat pumps exhibit distinct electricity consumption patterns relative to buildings heated primarily by natural gas. Fig. 3 illustrates these differences using predicted CDF values from our trained MLP-ZIG electricity demand model, binned by outdoor temperature. Heat pump buildings show systematically higher CDF values in colder temperature bins, reflecting increased electricity consumption during heating season relative to the model's learned distribution. This observation motivates our classification experiments described in Section 4.5.

This visual separation translates into strong classification performance. Using CDF based features extracted from the MLP-ZIG model, logistic regression achieves an ROC-AUC of 0.77 on the held-out test

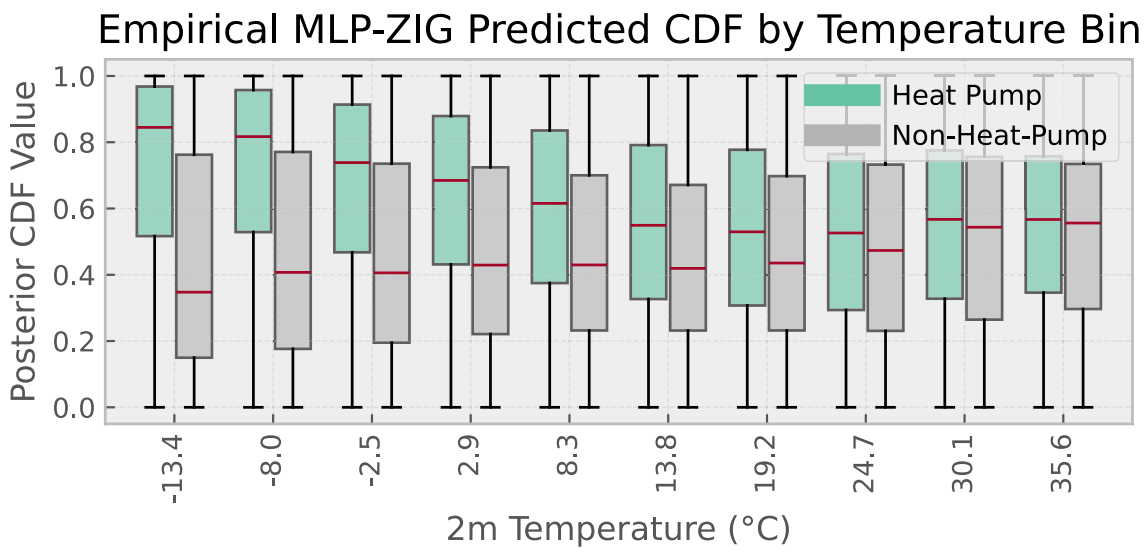

**Fig. 3.** Predicted CDF distributions by temperature bin for heat pump buildings versus gas heated buildings. Heat pump buildings show systematically higher CDF values at lower temperatures, reflecting increased electricity consumption during heating season.

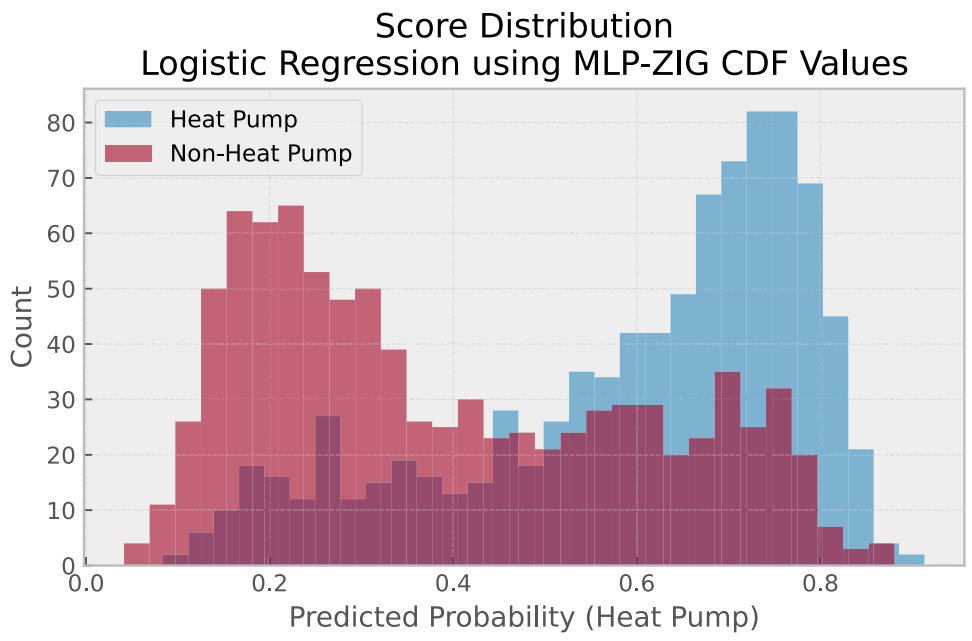

(a) CDF based features from MLP-ZIG electricity demand model.

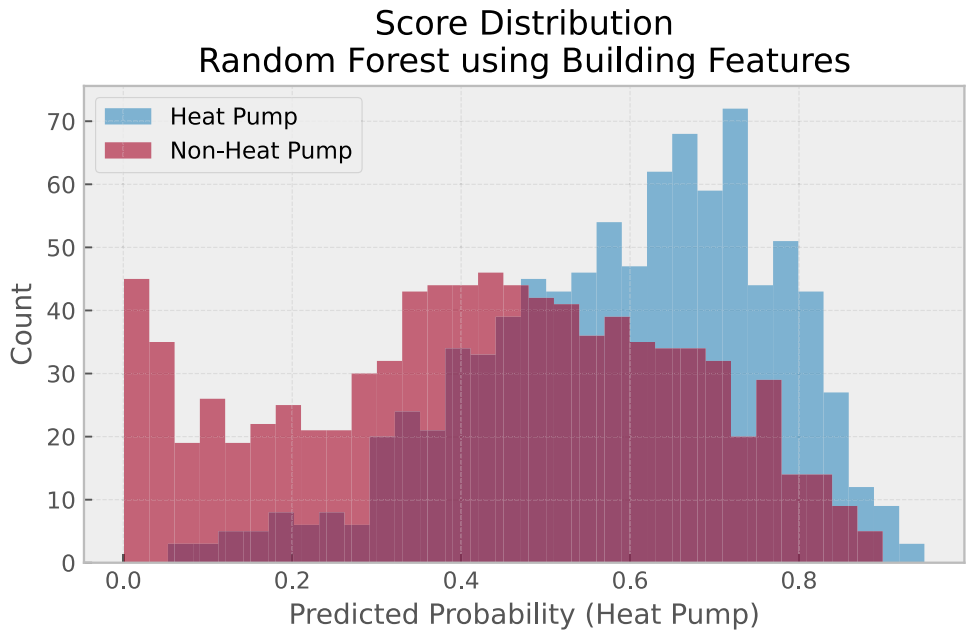

(b) Static building attributes from remote sensing and census data.

**Fig. 4.** Distribution of predicted heat pump probabilities for two classification approaches, illustrating separation between heat pump and non-heat pump buildings. Higher separation indicates better classification performance.

set. Using building features alone, random forest classification achieves an ROC-AUC of 0.72. Fig. 4 shows the distribution of predicted heat pump probabilities for both classification approaches.

We believe these results understate true model performance due to label noise in our ground truth data. Some buildings labeled as not having heat pumps may actually have heat pumps installed outside the rebate program, which would cause the classifier to be penalized for correct predictions. The building feature classifier is particularly valuable because it operates entirely on remote sensing and census data without requiring historical consumption records. Full classification metrics are provided in Supplementary Section S9.

## 6. Discussion

In this section, we evaluate our models' performance relative to the research community-standard ResStock framework, highlighting key advancements and areas for improvement. Our results indicate that machine learning-based demand forecasting, leveraging high-resolution geospatial and remote sensing features, outperforms traditional physics-based simulation models by providing more accurate and granular predictions. We also emphasize improvements in capturing demand heterogeneity and probabilistic uncertainty.

Beyond modeling improvements, we explore broader applications of our framework. Our high-resolution demand forecasts offer actionable insights for grid planning, electrification strategies, and policy development. By enabling more precise estimation of heating electrification impacts, characterizing climate-induced demand shifts, and identifying disparities in heat pump adoption, our work provides a scalable and data-driven approach to inform decarbonization pathways.

### 6.1. *State-of-the-art model performance*

Our models significantly outperform the ResStock baseline for both heating and electricity demand forecasting, demonstrating the advantages of a machine learning-driven approach that leverages remote sensing data at the building level. As detailed in Tables 1 and 2, our models achieve substantial error reductions relative to ResStock. For heating demand, our MLP-ZIG model reduces RMSE by 18.8% and WIS by 59% compared to the ResStock baseline. For electricity demand, our reduced-feature MLP-ZIG model reduces RMSE by 27.6% and WIS by 59%.

The superior performance of our models can be attributed to their ability to learn complex relationships between building characteristics, weather conditions, and energy consumption patterns. These relationships are difficult to encapsulate through traditional physics-based simulations. ResStock, for example, relies on predefined building archetypes and requires highly detailed input parameters such as insulation levels, HVAC system specifications, and occupancy schedules. However, obtaining such high-resolution data at scale is impractical, as no organization collects comprehensive building-level information at this level of detail across diverse regions.

The similar performance of the MLP-ZIG model, the reduced-feature MLP-ZIG variant, and the Gradient Boosting and Random Forest baselines suggest that the remote sensing feature engineering framework, rather than the neural network architecture alone, drives the observed improvements. That scalable, openly available geospatial data can support high-accuracy demand forecasting across multiple model architectures. Our results underscore the strength of the ML paradigm for our heating and electricity demand forecasting applications; they suggest that even greater accuracy may be achieved with additional features such as raw satellite imagery, model optimizations, and more labeled consumption data.

The strong performance of our reduced-feature baseline also highlights feature selection as a promising direction for future research. Automated feature selection for high-dimensional predictive modeling remains an active area of investigation, with recent advances demonstrating effectiveness across diverse domains [44–48]. While we employed SHAP-based importance rankings to construct our reduced-feature baseline, future work could explore these and other automated selection methods to further optimize the trade-off between model complexity and predictive accuracy.

#### 6.1.1. *Comparing our models for electricity and heating demand*

Modeling electricity demand proves easier than heating demand for our ML models, reflected by substantial error metric improvements relative to our ResStock baselines and notably stronger calibration as measured by the Kolmogorov–Smirnov statistic. Several factors contribute to this superior performance.

Electricity demand rarely drops to zero. Even during overnight hours when human activity is minimal, electricity consumption persists due to baseline loads including refrigerators, HVAC systems, and standby electronics. The diverse end uses of electricity contribute to smoother consumption patterns: space cooling, lighting, electronics, and appliances follow cyclical or seasonally predictable trends, allowing our models to effectively average out fluctuations. This inherent stability in consumption behavior enhances the accuracy and calibration of electricity demand forecasts.

In contrast, heating demand forecasting faces fundamental challenges when using natural gas consumption as a proxy. Several non-heating gas end uses introduce noise into the training signal. Cooking loads produce sharp, unpredictable spikes that create frequent outliers in the consumption distribution, penalizing model predictions even when the underlying heating demand is correctly estimated. Pilot lights, dryers, fireplaces, and pool heaters contribute intermittent baseline consumption that varies across households in ways that are difficult to predict from exogenous features. These secondary loads blur the boundary between zero and positive consumption that the ZIG distribution assumes, causing the model to assign nonzero probability mass to heating demand during periods when only cooking or other end uses occur. Despite these calibration challenges, the heating model still achieves substantial improvements over baselines on point and probabilistic accuracy metrics, demonstrating that even imperfectly calibrated probabilistic forecasts provide valuable uncertainty quantification.

### 6.2. Aggregate-level planning implications

The cluster-level results presented in Section 5.2 have important implications for grid planning at aggregated spatial scales such as feeders, districts, and balancing areas. The divergence between staggered and time-aligned aggregation modes indicates that weather data quality is the binding constraint on aggregate-level forecast accuracy. The ERA5 reanalysis data used in this study have 0.25° spatial resolution, which introduces systematic biases relative to local conditions that do not cancel when buildings share the same weather inputs. Higher-resolution weather data from commercial providers or bias-corrected numerical weather prediction could substantially reduce these errors and bring time-aligned accuracy closer to the staggered baseline.

For long-term capacity planning, where forecasts are often averaged over scenarios or seasons, our models can achieve very low aggregate errors, reliably informing infrastructure sizing decisions. For near-term operational forecasting, weather data quality remains the primary target for improvement. Detailed methodology and extended planning implications are provided in Supplementary Section S7.3.

### 6.3. Heat pump classification and propensity modeling

The classification results presented in Section 5.6 demonstrate that our framework captures meaningful signal about heating system type. The building feature classifier is particularly valuable because it operates entirely on remote sensing and census data without requiring historical consumption records. This independence from consumption data enables two complementary applications for electrification planning.

First, by identifying buildings likely to have heat pumps using only remote sensing data, utilities can assess current heat pump penetration across their service territory without requiring detailed customer surveys. Second, the same model can be interpreted as a heat pump propensity model that identifies buildings likely to adopt heat pumps in the future. Buildings with high predicted probabilities that do not currently have heat pumps represent early adoption candidates, while buildings with lower probabilities may require additional incentives or targeted outreach.

These capabilities support scenario analysis for electrification planning. For example, if policymakers anticipate a 20% increase in heat pump adoption due to a new incentive program, the propensity model can identify which neighborhoods and building types are likely to electrify first based on their feature profiles. Conversely, the model can highlight areas with low predicted adoption likelihood where additional intervention may be needed to achieve equitable electrification outcomes. By revealing the spatial distribution of current adoption and future propensity, this framework can help utilities and policymakers target incentives, infrastructure investments, and outreach programs toward achieving their heating decarbonization objectives.

### 6.4. Broader applicability and deployment

A key advantage of our remote sensing feature engineering framework is its transferability beyond residential applications. Because the approach relies on openly available geospatial data rather than building-specific parameters, it can be extended to other building types where consumption data are available for training. Preliminary experiments applying our methodology to commercial buildings demonstrate this transferability, achieving reasonable accuracy across 14 NAICS categories for electricity and 6 categories for natural gas despite substantially smaller training samples than our residential models. While data volume constraints prevent comprehensive commercial analysis in this study, these initial results suggest that the same geospatial features that drive residential demand prediction carry meaningful signal for commercial buildings as well. Detailed methodology and results are provided in Supplementary Section S10.

Beyond transferability, our approach offers additional advantages related to characterizing demand heterogeneity, quantifying uncertainty, and complementing physics-based models like ResStock. These capabilities have important implications for grid planning, policy development, electrification scenarios, and climate-resilient infrastructure planning. We note that our models are trained on the U.S. Mid-Atlantic region; deploying elsewhere would require regional retraining or transfer learning approaches, though the globally available nature of our input features facilitates such extensions. We also address practical deployment considerations including data governance, workflow integration, and computational scaling in Supplementary Material Section S11.

## 7. Conclusion

Our work presents a novel, high-resolution, probabilistic framework for forecasting residential heating and electricity demand, addressing critical limitations in existing modeling approaches. By integrating machine learning techniques with multimodal geospatial data, we provide more accurate, granular, and scalable demand estimates that outperform industry-standard models such as NREL's ResStock. Our probabilistic approach not only enhances predictive accuracy but also introduces uncertainty quantification, a critical component for informed decision-making in energy planning and policy.

Our results demonstrate notable improvements over traditional physics-based simulations, particularly in capturing real-world demand heterogeneity. The ability to model demand at the building level allows for more precise infrastructure planning, ensuring that utilities, policymakers, and regional planners can optimize resource allocation, anticipate grid constraints, and design policies that reflect actual energy consumption patterns.

The implications of our findings extend beyond forecasting accuracy. Our framework enables heat pump classification using learned baseline consumption patterns and also by using building attributes alone. This capability allows utilities to assess existing heat pump penetration and identify buildings amenable to electrification without detailed customer surveys. Our model provides valuable insights for electrification strategies, allowing policymakers to evaluate the grid impacts of heat pump adoption and develop targeted incentive programs that maximize decarbonization benefits. Additionally, the probabilistic nature of our approach equips decision-makers with the tools to assess risks and uncertainties, fostering more adaptive energy planning strategies.

By making our framework open-source and scalable, we contribute to the growing body of research supporting data-driven energy system modeling. Our work lays the foundation for further advancements in demand forecasting, including refinement of predictive capabilities through additional data sources. As the transition toward electrified heating accelerates, our model has the potential to serve as a critical tool for optimizing investments, improving grid resilience, and advancing global decarbonization efforts.

## CRediT authorship contribution statement

**Stephen J. Lee:** Writing – review & editing, Writing – original draft, Visualization, Validation, Supervision, Software, Resources, Project administration, Methodology, Investigation, Funding acquisition, Formal analysis, Data curation, Conceptualization. **Cailinn Drouin:** Writing – review & editing, Writing – original draft, Validation, Resources, Methodology, Investigation, Formal analysis.

## Declaration of Generative AI and AI-assisted technologies in the writing process

During the preparation of this manuscript, the authors used Claude by Anthropic to improve the language, grammar, and clarity of the text. Claude and Codex by OpenAI were used for routine coding tasks such as debugging and syntax. The authors reviewed all AI-assisted text and code, and take full responsibility for the integrity of the research and the accuracy of all content.

## Declaration of competing interest

The authors declare that they have no known competing financial interests or personal relationships that could have appeared to influence the work reported in this paper.

## Acknowledgments

We are grateful to our utility partners and the MIT Energy Initiative Future Energy Systems Center for their support and contributions to our research. We specifically want to thank Christopher Knittel, Dharik Mallapragada, Randall Field, and Raanan Miller for facilitating the programs and partnerships that made this research possible. We are also grateful to the anonymous reviewers whose constructive comments helped strengthen this manuscript.

## Appendix A. Supplementary data

Supplementary material related to this article can be found online at https://doi.org/10.1016/j.egyai.2026.100726.

## Code and data availability

The trained models and code for generating building-level heating and electricity demand estimates and forecasts are released as open-source software under the MIT License at https://github.com/stephenjlee/mlp-zig-heat-elec-demand.